# COMPLETE

## A flagship mission for complete understanding of 3D coronal magnetic energy release

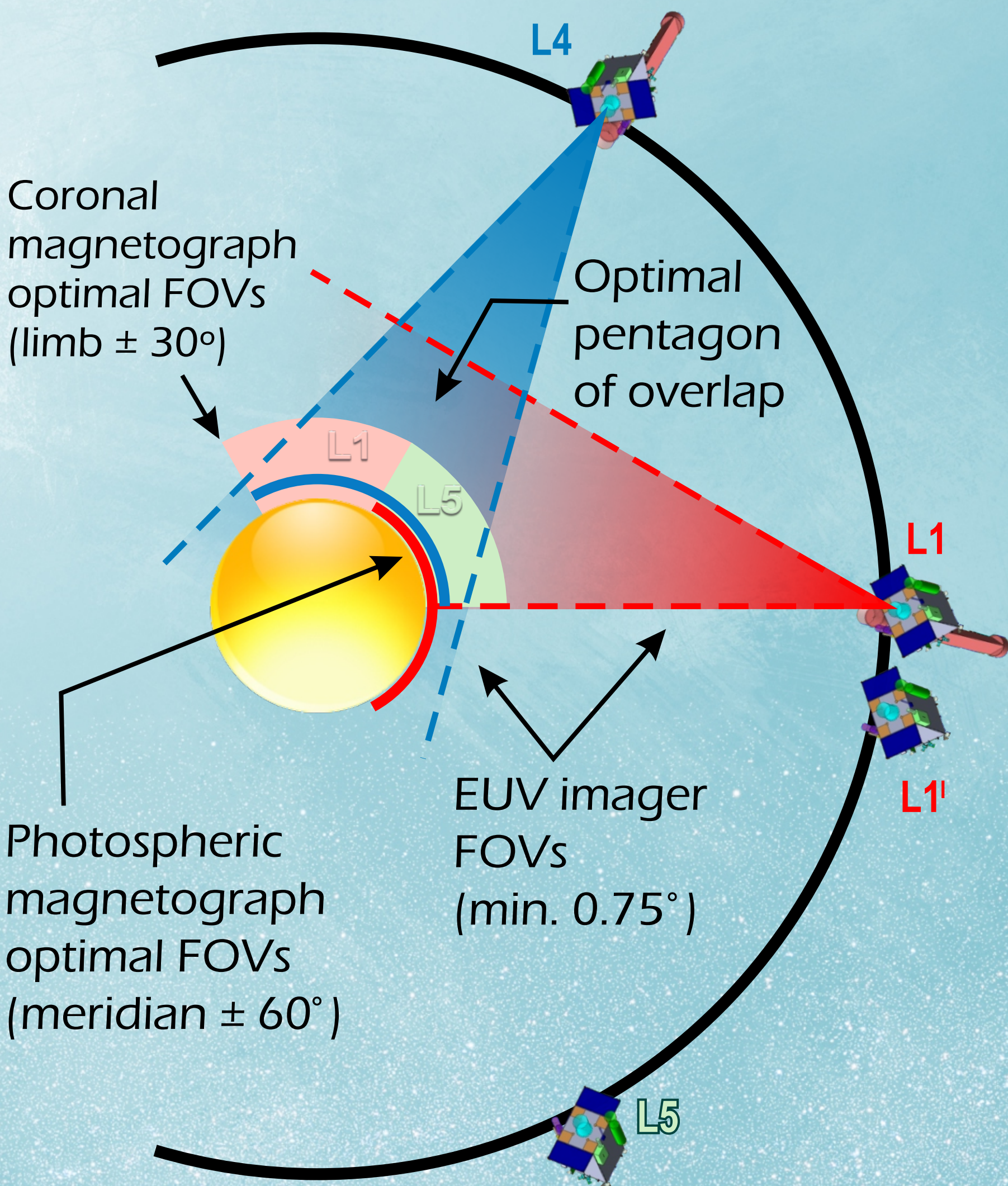

**Synopsis:**

COMPLETE is a flagship mission concept combining broadband spectroscopic imaging and comprehensive magnetography from multiple viewpoints around the Sun to enable tomographic reconstruction of 3D coronal magnetic fields and associated dynamic plasma properties, which provide direct diagnostics of energy release. COMPLETE re-imagines the paradigm for solar remote-sensing observations through purposefully co-optimized detectors distributed on multiple spacecraft that operate as a single observatory, linked by a comprehensive data/model assimilation strategy to unify individual observations into a single physical framework. We describe COMPLETE's science goals, instruments, and mission implementation. With targeted investment by NASA, COMPLETE is feasible for launch in 2032 to observe around the maximum of Solar Cycle 26.

Amir Caspi[1], Daniel B. Seaton[1], Roberto Casini[2]; Cooper Downs[3], Sarah Gibson[2], Holly Gilbert[2], Lindsay Glesener[4], Silvina Guidoni[5], Marcus Hughes[1], David McKenzie[6], Joseph Plowman[1], Katharine Reeves[7], Pascal Saint-Hilaire[8], Albert Y. Shih[9], Matthew J. West[1]

[1]Southwest Research Institute, [2]NCAR High Altitude Observatory, [3]Predictive Sciences Inc., [4]University of Minnesota, Twin Cities, [5]American University, [6]NASA Marshall Space Flight Center, [7]Center for Astrophysics | Harvard & Smithsonian; [8]Space Sciences Laboraotry, University of California, Berkeley, [9]NASA Goddard Space Flight Center

## 1 Science Goals

The solar corona is a dynamic three-dimensional environment whose physics is dominated by its complex magnetic field. Transformative progress in understanding how magnetic energy powers energetic processes in the corona – solar eruptions, active region heating, solar wind acceleration, and more – requires measuring the coronal magnetic field and determining its 3D structure, neither of which are within current capabilities. Furthermore, recent analyses have cast doubt on previously used 3D reconstruction techniques, suggesting that 3D structure cannot be accurately reconstructed solely using measurements of optically thin emission, without direct knowledge of the underlying magnetic field.

Resolving this dilemma lies within present-day technical capabilities, with sufficient investment. **COMPLETE combines comprehensive magnetography with broadband spectroscopic imaging from multiple viewpoints to enable tomographic reconstruction of 3D coronal magnetic fields and associated dynamic plasma properties that provide direct diagnostics of energy release**. COMPLETE achieves this through highly co-optimized instrument suites coupled with an integrated data assimilation strategy that unifies its deliberately complementary measurements and data-constrained modeling within a unified physical framework.

COMPLETE implements a grand vision to overcome the challenges imposed by the broad range of observational regimes that must be sampled to completely characterize the corona and answer long-standing questions in solar physics. These challenges require a mindful strategy for integrating such widely disparate data that cannot be achieved through piecemeal solutions. COMPLETE embraces a dramatic paradigm shift from prior mission implementation, with multiple detectors distributed at different vantage points but purposefully targeted and co-optimized to work together as a single observatory. **COMPLETE adopts a policy of focused innovation, extending key principles of inventiveness end-to-end across all mission aspects, from instrumentation to communications, to data processing and end-user analysis tools.**

COMPLETE's comprehensive science goal is to understand the magnetic energy storage and release that drives all heating and dynamics in the corona. Its four primary science questions to address this goal are enumerated in the Science Traceability Matrix (Fig. 1) and are based on the detailed science discussion in Caspi et al. (2022).

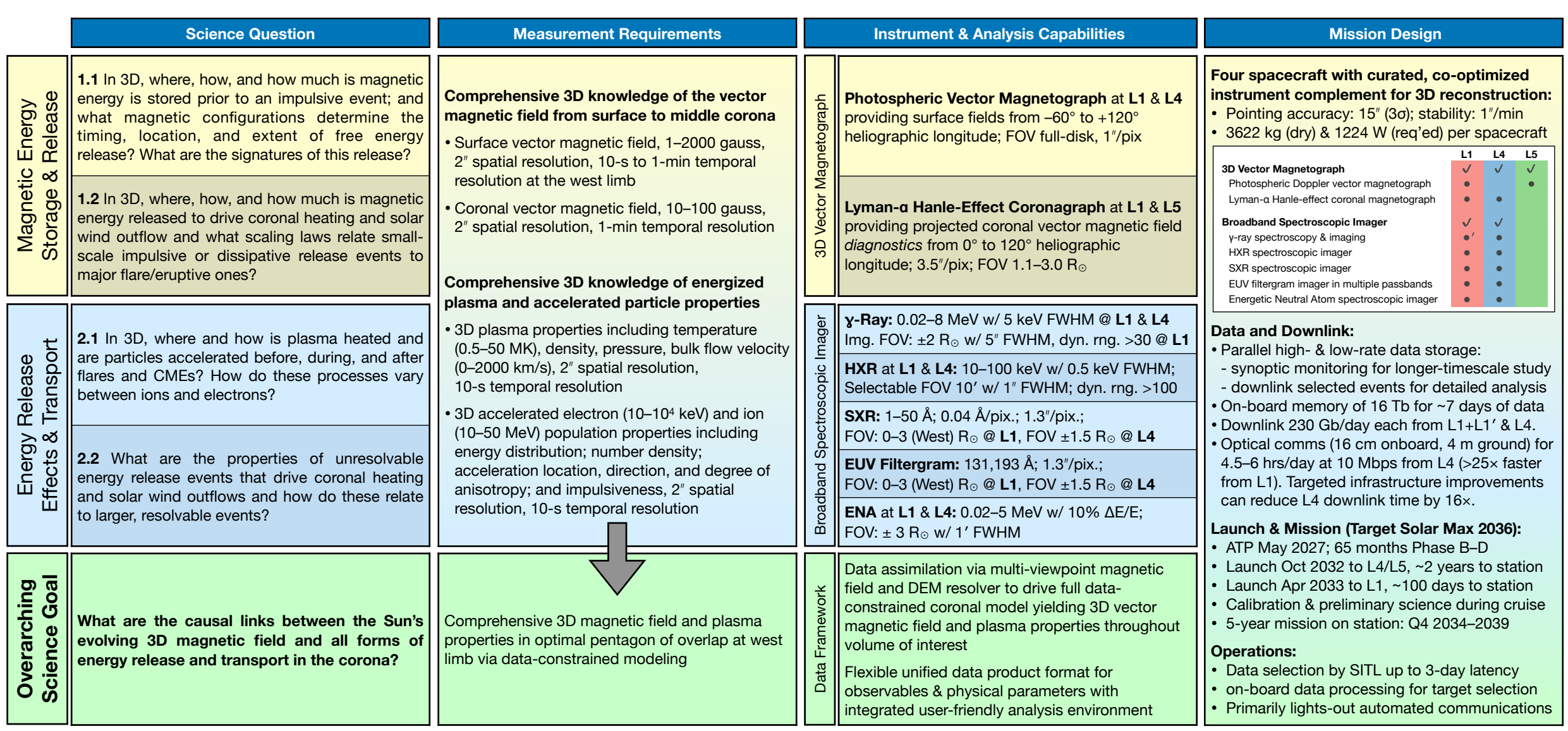

Fig. 1. A condensed version of the COMPLETE science traceability matrix highlights the physical parameters and associated measurements required to close the science questions. The COMPLETE baseline implementation is wholly feasible based on current and planned technological roadmaps.

## 2 Investigation Description

The COMPLETE mission architecture provides a simple, robust, high-heritage design in familiar orbits to maximize science return. Deployment of four COMPLETE observatories, each with a tailored complement of instruments and a high-speed Optical Laser Communications system, into halo orbits around the Earth-Sun Lagrange points L1, L4 and L5 provides an implementation approach that enables breakthrough Heliophysics systems science and data processing.

The COMPLETE team used the NASA GSFC Mission Design Lab (MDL) to formulate a mission concept centered on the COMPLETE L4 observatory, which would drive out the necessary design work for the entire COMPLETE mission architecture. The MDL was directed to assume all four COMPLETE observatories would use the same L4 mission spacecraft bus when estimating costs for the entire COMPLETE mission. The sections that follow highlight key aspects of the COMPLETE mission implementation, as formulated by the MDL and science team.

### 2.1 Mission Overview

The COMPLETE mission is designed to provide comprehensive system-level integration of its science objectives. The mission architecture and design achieve all objectives with four total observatories: one each in orbits around Lagrange points L4 and 5, and two at L1 (*see cover page image*). These "fixed" vantage points allow for continuous observation of the Sun throughout the 5-year prime mission. Offsetting the L1 fields of view provides focus on the West limb. Launch in late 2032 enables observations around the peak of Solar Cycle 26. COMPLETE is a Class C risk classification mission with a 7-year design life (2 years transit, 5 years Phase E). The L4 and L5 observatories will be launched from KSC in December 2032 on a single Falcon Heavy, followed 6 months later by the launch of the L1 and L1′ observatories on a second Falcon Heavy.

### 2.2 Instruments

COMPLETE implements two targeted instrument suites –a broadband spectroscopic imager and a comprehensive 3D vector magnetograph – distributed across its multi-viewpoint spacecraft. To accommodate the realities of measurement physics, these suites comprise multiple detectors, each optimized both for its specific observations and to function as an element of a single integrated observatory. The instrument requirements are specifically crafted to enable seamless integration within COMPLETE's data assimilation and physical modeling framework to generate unified data products containing the meaningful physical properties of the underlying coronal plasma (see Caspi et al. 2022, Seaton et al. 2022). Predicted instrument performances meet measurement requirements and all instrument designs are derived from previously formulated or flown instruments, with appropriate engineering modifications required for COMPLETE.

#### 2.2.1 Broadband spectroscopic imager

Signatures of energy release cause emission spanning the entire electromagnetic spectrum. Emission in extreme ultraviolet (EUV), X-rays, and γ-rays provide the optimal diagnostics for plasma properties to characterize plasma heating, particle acceleration, and bulk flows. Each wavelength range requires different optics and detector technology, so COMPLETE implements this broadband suite as four co-optimized telescopes: an EUV filtergram imager in multiple relevant passbands; a soft X-ray (SXR) slitless imaging spectrograph; a hard X-ray (HXR) focusing-optics photon-counting spectroscopic imager; and a γ-ray Fourier indirect spectroscopic imager. An additional telescope measuring energetic neutral atoms (ENAs) provides multi-messenger remote sensing complementary to γ-rays. These telescopes are distributed across COMPLETE's multi-perspective spacecraft with optimized capabilities and fields of view to operate as a single observatory and enable 3D stereoscopic reconstruction of emitting coronal structures (e.g., Plowman 2021, 2022). Fig. 2 shows a condensed representation of simultaneous observations for a flare and CME.

**EUV filtergram imager:** The EUV Imager is based on the SDO/AIA and GOES/SUVI designs, and provides EUV imaging in two passbands, optimized for flare and global coronal temperatures

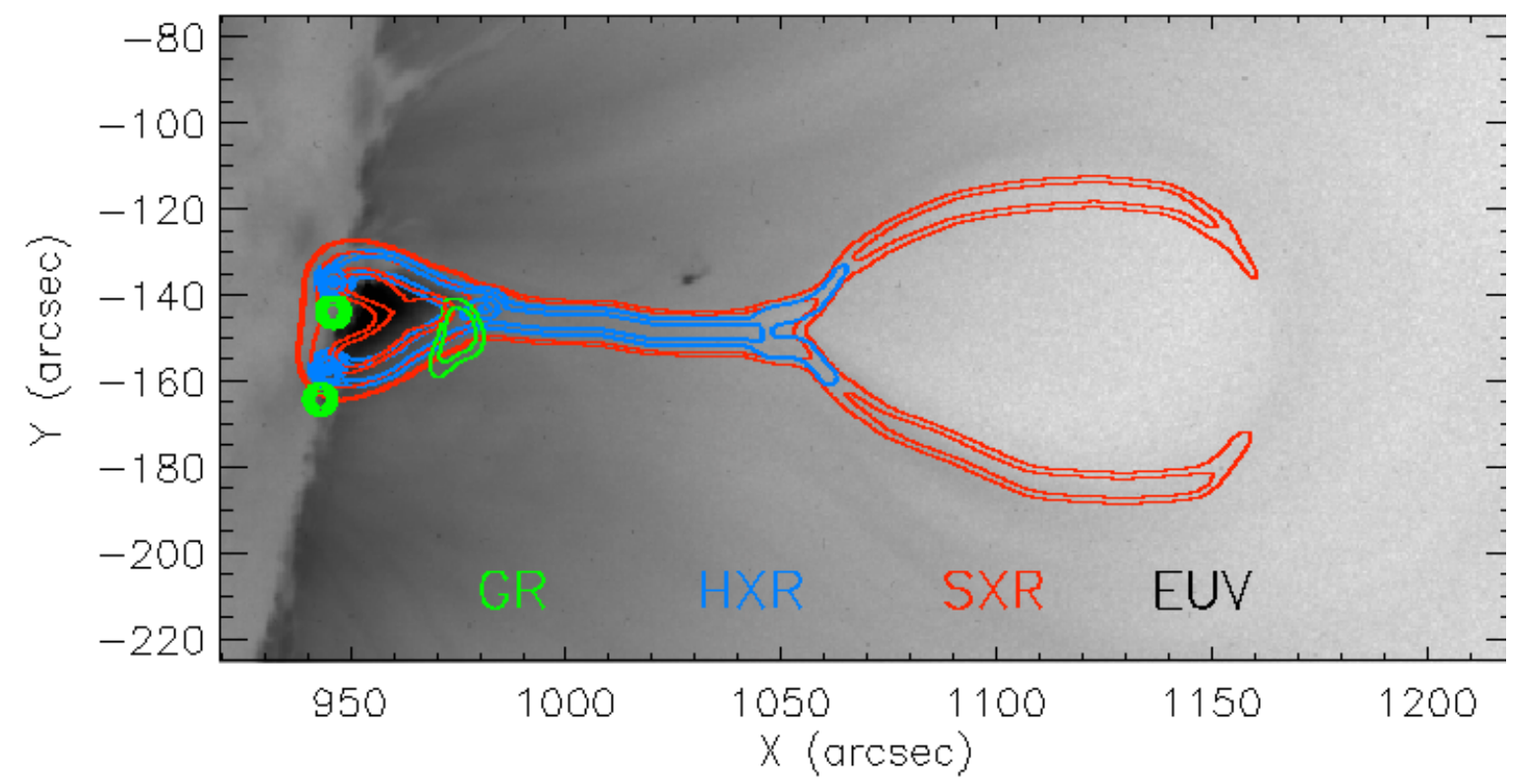

Fig. 2. Representations of COMPLETE's broadband spectroscopic observations for an eruptive flare on the West limb (as seen from L1), highlighting the diagnostics from each detector within the instrument. EUV and X-rays provide diagnostics of hot plasma while X-rays and γ-rays probe accelerated electrons and ions, both from the flare and CME. ENA emission from the CME shock front is at higher altitudes and not shown.

(131 and 193 Å passbands, respectively). The telescope is a Ritchey-Chrétien design, modified to flatten the focus across the field. The primary and secondary mirrors are each coated with Silicon/Molybdenum (Si/Mo), which has extensive heritage and demonstrated long-term stability. The channels are selected using focal plane filters (Al for 193 Å, Zr for 131 Å). The field of view is 0.75° with 1.3″/pixel. The FOV from L1 will be off-pointed so that the image is centered on the West limb of the Sun, whereas the imager at L4 will be pointed at disk center from that vantage point.

**SXR spectroscopic imager:** The SXR imager is a slitless spectrograph with focusing optics and transmission diffraction grating, leveraging heritage from Hinode/XRT, CubIXSS/MOXSI, and Chandra/HETG to provide measurements of coronal temperatures from ~1 to >30 MK and abundances of key trace elements (Fe, Ca, Mg, Si, S, Ne, O, Ar, C) to test models of plasma heating in flares and active regions. A 2k×4k detector provides a 0.75°×1.5° FOV with 1.3″/pix plate scale, matched to the EUV imager. The diffraction grating disperses X-rays from each source on the full Sun onto the detector, with dispersion in the 4k pixel direction, providing spectral diagnostics from <4 Å to >55 Å with 0.04 Å/pix plate scale from every pixel on the Sun simultaneously. Even-order dispersion is suppressed, and 3rd order provides even higher spectral resolution around critical spectral lines. The high quantum yield of X-ray photons enables thresholding at 1 Hz to reject "empty" pixels; faster readout at >100 Hz (e.g., like the PhoENiX detector on the FOXSI-3 rocket) would also enable direct photon counting to further improve discrimination between dispersed orders.

**HXR spectroscopic imager:** The HXR imager is based on the Focusing Optics X-ray Solar Imager (FOXSI; Christe et al. 2022) design proposed for Explorer-class missions. FOXSI uses electroformed Ni mirrors to directly focus HXRs above 3 keV. The FOXSI-4 sounding rocket is implementing design improvements to achieve 2.5″ FWHM / 10″ HPD, and additional engineering development by the time of COMPLETE PDR is expected to yield 1″ FWHM / 5″ HPD as required for COMPLETE. FOXSI's direct focusing imaging obtains orders-of-magnitude better sensitivity and dynamic range than previously available from indirect imaging. Multiple HXR sources within a flare can be easily separated in space, time, and energy, and coronal sites of electron acceleration can be systematically measured even in the presence of bright footpoint sources. Directly focusing HXRs requires small grazing angles and therefore a large separation (10–20 m) between the mirrors and detectors. The HXR imager uses a 14 m boom, proven feasible via designed and full analysis under the FOXSI SMEX Phase A concept study. CdTe detectors with rapid readout have also been demonstrated, and COMPLETE leverages this extensive prior design heritage.

**γ-ray imager and spectrometer:** The L1′ spacecraft carries a γ-ray spectroscopic imager based on the Gamma-Ray Imaging, Polarimetry, and Spectroscopy (GRIPS) instrument (Duncan et al. 2016). GRIPS provides a near-optimal combination of high-resolution imaging, spectroscopy, and polarimetry of solar-flare gamma-ray/hard X-ray emissions from 10 keV to 10 MeV. The key technology is that of a 3D position-sensitive Ge detector (3D-GeD) that allows the position and energy deposition of every photon interaction to be recorded individually, even within the same detector. GRIPS pairs the 3D-GeDs with a new single-grid Multi-Pitch Rotating Modulator (MPRM) imaging system mounted on a 20-m boom to provide a near-ideal point response function (5″ FWHM, full-Sun FOV) with 2× throughput per $cm^2$ of detector area compared to RHESSI. Compton scattering of incident photons (the dominant interaction in Ge at >150 keV) is tracked to reject background for spectroscopy and imaging, and also provides polarization information.

For mass purposes, the L4 spacecraft carries only the spectrometer portion of GRIPS, without the imaging grids and boom. Because the L4 craft will observe Earth-view West-limb eruptions from "above," its primary observations would be of footpoints, and hence spectroscopy captures the primary relevant measurements even without imaging.

**ENA spectroscopic imager:** The ENA Spectroscopic Imager (ENASI) meets the challenges specific to observing solar ENAs vs. observing ENAs in other contexts (e.g., from Earth's magnetosphere or from the heliopause). Solar-ENA measurements require both high effective area and high angular resolution, so the implementation approach is akin to an X-ray/γ-ray spectroscopic imager instead of a traditional ENA instrument. ENASI provides measurements from 20 keV to 5 MeV with ~500 $cm^2$ effective area and ~1′ angular resolution. The instrument consists of a 50%-throughput mask placed ~2 m in front of an array of pixelated Si detectors. No technology development is required at the component level.

ENASI achieves indirect imaging through the combination of the mask pattern and the position sensitivity of the detector (similar to GRIPS). This approach maximizes the imaging effective area at the expense of being able to meaningfully constrain the origin of an individual ENA. Each observed ENA can be "back-projected" to the mask pattern on the sky. ENA images are reconstructed by determining the source configuration consistent with a collection of observed ENAs.

### *2.2.2 Comprehensive 3D magnetograph*

Building truly constrained coronal models requires measurements of both the surface and coronal magnetic field, achieved through two instrument subsystems – a photospheric Doppler vector magnetograph and a Lyman-α Hanle-effect coronal magnetograph – that leverage overlapping lines of sight through COMPLETE's multi-perspective mission architecture. They are complemented by diagnostics of coronal structures from the Broadband Spectroscopic Imager suite. Combining these measurements to infer the 3D magnetic field is a global optimization problem: given magnetically sensitive coronal observations, determine the magnetic field distribution that generates them. Community tools like FORWARD, which synthesize coronal observables based on model/simulation input (Gibson et al. 2016), pull these strands together and may be used by inversion frameworks along with observations and global MHD simulations (e.g., Mikic et al. 2018) to solve for an optimized coronal magnetic field (e.g., Kramar et al., 2006, 2013, 2016; Dalmasse et al. 2019). COMPLETE provides powerful new information about the global 3D field with its multi-vantage measurements. *Even without a full 3D reconstruction, COMPLETE's quadrupolar vantage immediately diagnoses magnetic energy buildup in Earth-directed CME precursors using the coronal magnetograph* (Caspi et al. 2022; Casini et al. 2022; Gibson et al. 2022).

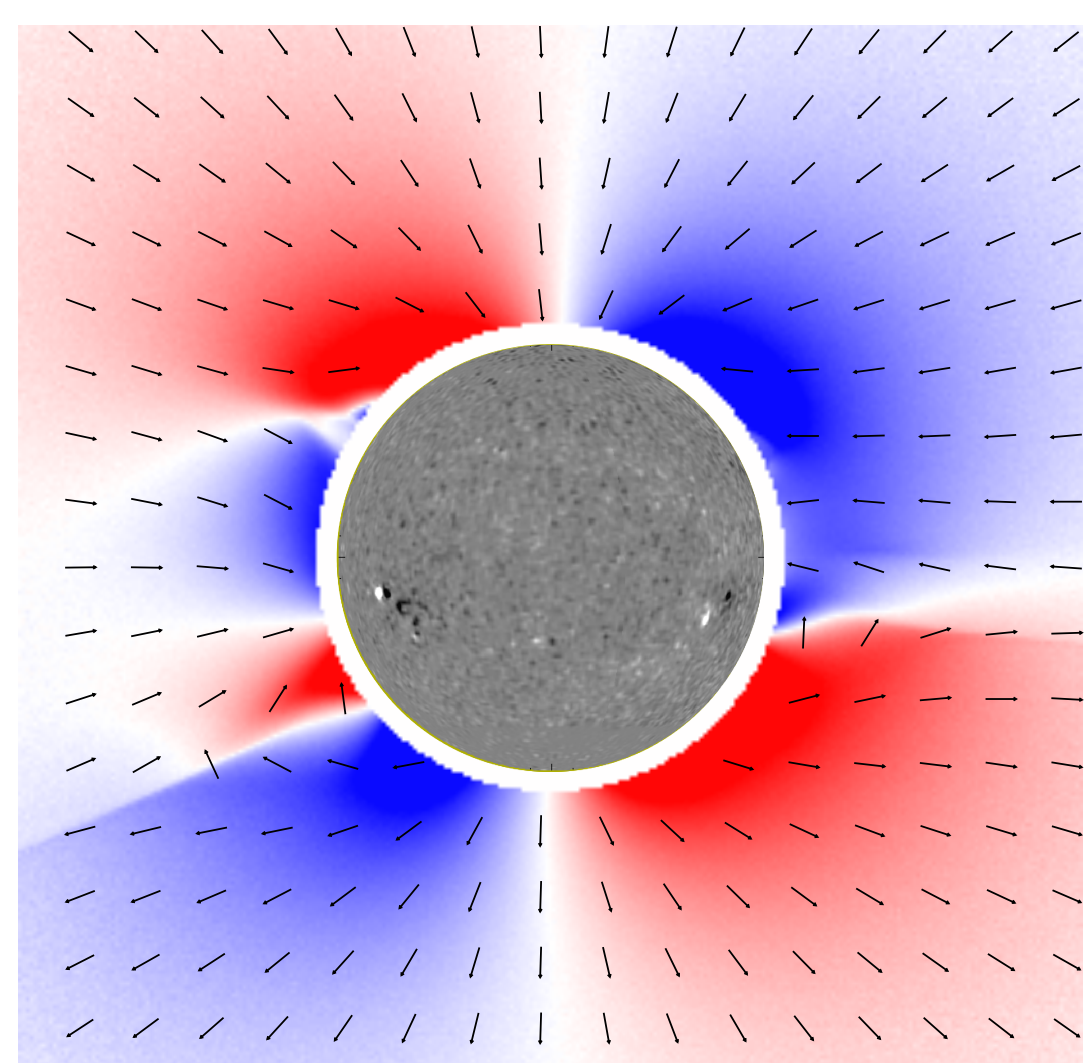

Fig. 3. Simulated composite observations from COMPLETE's comprehensive 3D magnetograph, showing photospheric and coronal measurements. The Lyman-α polarization signal is diagnostic of the line-of-sight integrated coronal vector field.

**Photospheric magnetograph:** The requirements are directly traceable to COMPLETE's core concept: the magnetic measurements must have adequate spatial resolution and magnetic sensitivity (via polarization sensitivity) to provide robust measurements of the vector magnetic field and current density at the base of pre-eruption coronal structures, sufficiently paired with the coronal measurements to enable a full 3D reconstruction of the magnetic field. The telemetry, mass, and volume constraints of the COMPLETE mission are challenging. The magnetograph considered for COMPLETE is based on the Kobayashi et al. (2020) design for a notional High-Inclination Solar Mission, which included an imaging spectropolarimenter comprising a telescope, a polarizer, and a narrowband tunable filter. The COMPLETE magnetograph baselines the Fe I line at 6173 Å, with ~1″ FWHM spatial resolution, and a full-disk field of view. If needed, an alternative concept designed for compactness – the Compact Doppler Magnetograph (CDM) designed for the Solaris MIDEX mission concept (Hassler et al. 2022) – could be modified for vector field measurements with additional engineering development.

**Lyman-α Hanle coronagraph:** Linear polarization of Ly-α resonant scattering emission is sensitive to magnetic field strengths of 5–200 G, ideally covering the expected range in closed-field regions of the low corona. Ly-α is the strongest solar UV emission, visible even high in the corona (see, e.g., recent observations from Solar Orbiter / METIS; Romoli et al. 2021). The higher brightness ratio between corona and disk in Ly-α greatly reduces the impact of instrumental stray-light for coronametry (see Casini et al. 2022 for details of the enabling polarimetric diagnostics and methodology). All modeling tools necessary for interpreting coronal Ly-α polarimetric observations are already implemented in the FORWARD code (Gibson et al. 2016, Zhao et al. 2019, 2021).

Thanks to the higher corona-to-disk brightness ratio in the FUV, and heritage mirror-polishing techniques and high-reflectivity coatings, the Ly-α coronal magnetometer adopts a fully reflective, internally occulted coronagraph. With such an instrument, coronal polarimetry up to ~3 $R_{sun}$ becomes possible with apertures of only ~0.1 m. An inverse-occulted coronagraph design is preferrable, as it allows a better control of thermal load and stray light. Additionally, such a design choice enables use of a quad-cell sensor behind the inverse occulter for real-time image-stability control, which is necessary for high-precision polarimetry. The instrument adopts a dual-beam polarimeter, consisting of a heritage $MgF_2$ $\lambda/2$ retarder for the polarization modulator, and a novel, high-contrast polarizing beam splitter for Ly-α (Gutiérrez-Luna et al. 2022). This design provides optimal linear-polarization efficiency, and implements the "beam swapping" technique, making the data practically insensitive to detector-gain and throughput imbalance between the two beams.

### *2.3 Spacecraft bus flight systems*

The COMPLETE flight system is designed to meet the mission requirements defined in the STM. Fig. 4 shows the L4 observatory as deployed and stowed, and is representative of the other three observatories that are built with identical buses to reduce cost and risk. Only minor tailoring is required to accommodate the unique instrument complement of each bus. *All elements of the COMPLETE bus and mission architecture are feasible with current or in-development technology.*

The COMPLETE observatory is 3-axis-stabilized, with 4 reaction wheels (RW), 2 solid state inertial reference units (SSIRU), a 2-headed star tracker system, and 12 coarse sun sensors. These provide 15″ (3σ) pointing accuracy and 1″/min stability as required by COMPLETE. The avionics is based on a heritage Moog Integrated Avionics Unit (IAU) and includes a 16 Tb solid state recorder. The communications subsystem includes a COTS laser optical terminal with a 2-axis gimbal for high-speed science data downlink, and an X-band RF system for launch and early-orbit (L&EO), ranging and command uplinks, and house-keeping telemetry downlink, via an omni antenna pair and medium gain antenna. Electrical power of >1400 W is supplied by 4.2 $m^2$ solar arrays and a 60 Ah Li-Ion battery. COTS parts are heavily leveraged for cost and risk reduction.

The bus structure is an aluminum rectangular frame with honeycomb panel closeouts. Thrust cylinders support 4 propellant tanks. Mechanisms deploy the 2 solar array panels. The observatory is a thermally passive design, cold biased, with thermal blankets, radiators and heaters. The propulsion system is bipropellant and uses one 100 lbf main engine, and eight 5 lbf ACS thrusters for orbit trajectory and insertion maneuvers, orbit station keeping, and RW momentum dumps. There is one fuel tank for the HPH, two spherical Ox tanks, and two spherical pressurant tanks. The propulsion system delivers 825 m/s delta-V. A monopropellant system is also feasible to reduce mass, complexity, and cost; a trade study is in process.

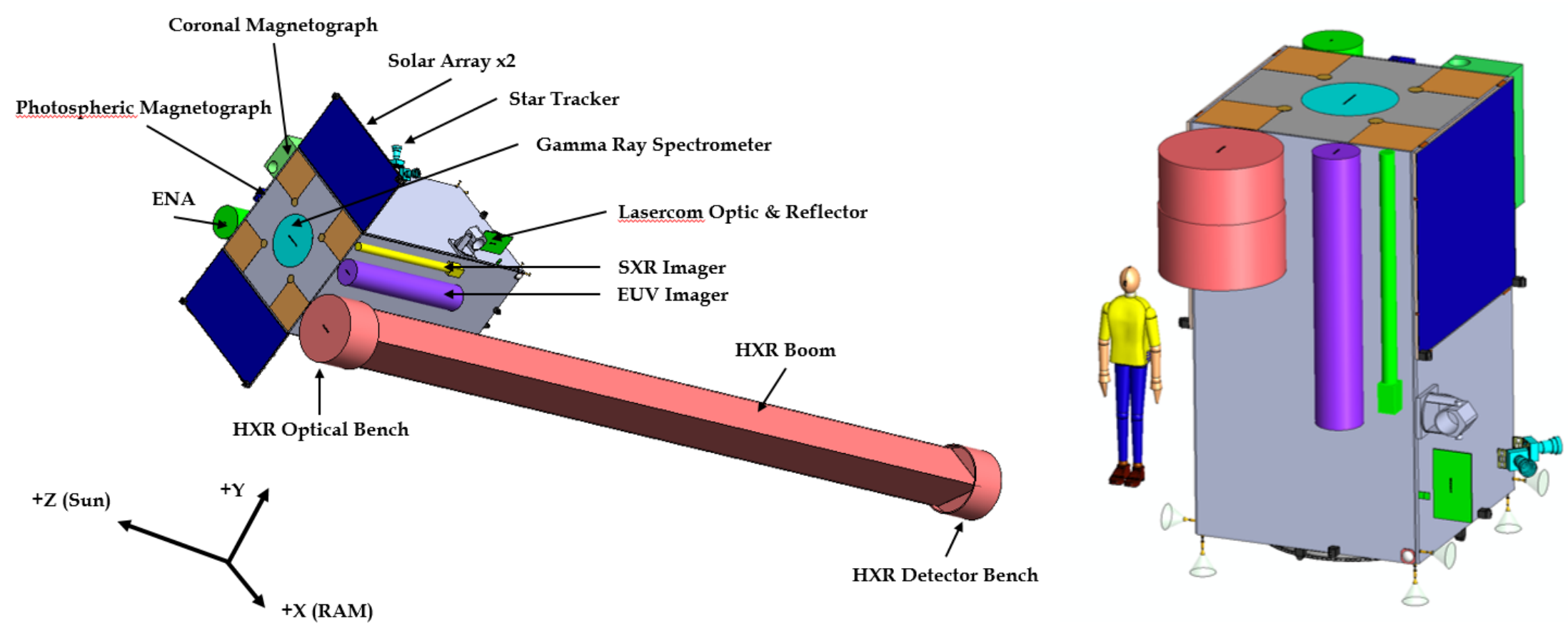

Fig. 4. COMPLETE L4 observatory in deployed and stored configurations. The L1, L1′, and L5 buses are identical except for instrument accommodations, to reduce cost and risk.

### *2.4 Concept of operations*

After launch of the L4/L5 Falcon Heavy, it takes ~2 years for all 4 observatories to be in position before beginning the 5-year Phase E. There are a handful of few post-launch maneuvers during the first 2 weeks, followed by insertion burns into orbit. There are no station-keeping maneuvers required for the L4 or L5 observatories. Each observatory will take 6 months post-launch for checkout and calibration to ensure all flight systems are functioning properly. Mission operations teams are staffed at different levels depending on the mission phase.

Instrument(s) are powered 24×7 during science mode, pointing at the Sun and taking data. The COMPLETE observatories use X-band for housekeeping and commanding, and ranging via daily Deep Space Network (DSN) contacts. High-speed science data is telemetered daily via the observatories Optical Laser Communications (OLC) subsystem to a JPL 4m terrestrial optical telescope (6 hrs average downlink time per day from L4 with existing infrastructure; the L1 data can be downlinked ~20× faster). The observatory executes simple 1–2× per year automated calibration maneuvers that offpoint from the Sun by a few degrees, including cruciform scans across the Sun and observations of the Crab nebula and pulsar.

With the exception of the planned but yet-to-be-funded 4-m JPL terrestrial optical telescope receiver, all COMPLETE ground system elements and Mission Operations Center (MOC) and Science Operations Center (SOC) elements are high heritage. Existing 4-m astronomical telescopes such as Palomar could be adapted for daytime optical communications if the JPL facility is delayed. Additional improvements to the optical infrastructure are discussed in §3.

### *2.5 Cost and schedule*

COMPLETE cost estimates were performed independently by both the GSFC MDL and by Technomics Inc. Both organizations used the NASA Instrument Cost Model (NICM) to arrive at 50% confidence level (CL) costs for the instruments; Technomics used the PCEC cost model for all other WBS cost estimates, including a 50% CL cost for WBS 6. The MDL used the SEER-H cost model to arrive at a 50% CL for WBS 6 and used a different percentage of WBS 5+6 to derive costs for the non-hardware WBSs. Launch Vehicle costs reflect two Falcon Heavy vehicles. Table 1 shows the Phase B–F mission costs, excluding technology maturation costs, estimated by each of these two organizations. Two independent organizations, whose estimates are ~10% apart from each other, adds credibility that these COMPLETE mission lifecycle costs are realistic given the mission and technical baseline. **We emphasize that a \$2.5B FY22 mission cost is equivalent to the inflation-adjusted cost for the SoHO flagship mission (\$1.27B FY95) and only 40% more than the inflation-adjusted cost for Parker Solar Probe (\$1.5B FY17) but with >2× the spacecraft and instruments.** ***COMPLETE's cost is thus entirely in-family for flagship-level missions and groundbreaking science return.*** To reduce financial impact to NASA, COMPLETE could also be implemented as a partnership with international agencies (e.g., ESA, JAXA, ISRO). A new program office is recommended to implement the strategic coordination required to realize COMPLETE's highly co-optimized architecture and integrated data assimilation framework.

**Table 1. COMPLETE Mission Costs FY22 $M**

| WBS Element | | Tech. | MDL |
|---|---|---|---|
| 1, 2, 3 | PM, Systems Eng., S&MA | 180 | 224 |
| 4 | Science | 84 | 97 |
| 5 | Payload Suite | 973 | 639 |
| 6 | Spacecraft | 239 | 1063 |
| 7 & 9 | Mission Ops. & Ground | 351 | 90 |
| 8 | LV Services | 323 | 323 |
| 10 | Systems I&T | 123 | 77 |
| | **Total** | **2191** | **2513** |

The COMPLETE team established a 71 mo. Phase B–D development plan and 5-year Phase E schedule, shown in Fig. 5. It is assumed that technology maturation activities to TRL 6 will already be completed prior to entering Phase A.

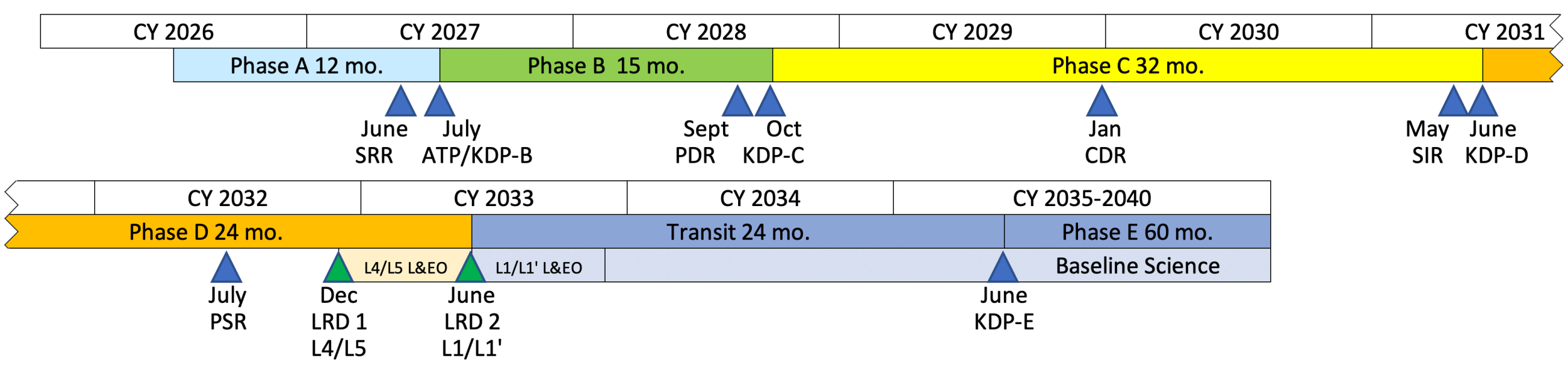

Fig. 5. COMPLETE development and mission schedule.

## 3 *Technology Development Needs*

### *3.1 Instrument development*

Each COMPLETE instrument leverages extensive current and prior development and missions. No specific technological development is required. We assume each instrument is at TRL 6 by mission PDR and do not explicitly include related engineering development costs in the mission cost above, but an estimated breakout is available in the full mission report upon request.

### *3.2 Optical communications infrastructure*

COMPLETE's extensive data generation and downlink requirements exceed current DSN RF downlink capacity by two orders of magnitude, necessitating OLC. COMPLETE's baseline needs can be met through existing and in-development technology, including an on-board terminal with 16-cm aperture and a ground receiver with 4-m aperture, with ~6 hours of downlink per day from L4 (much less from L1). Existing infrastructure can be used to meet these requirements. However, a 16× improvement in downlink speeds can be realized through modest improvements already planned for the roadmap: quad-band wavelength multiplexing, and an 8-m ground aperture. Multiplexing is already in development and an 8-m station is already planned for deployment by JPL around 2035. Additional targeted investments by NASA in these two areas, to hasten their schedule by a few years, would enable their use on COMPLETE and other deep-space missions to significantly reduce downlink times and thus resource competition, and/or to increase downlinked science data volume and thus the science return (Shelton et al. 2022). Multiplexing for the on-board terminal would need to be at TRL 6 by PDR (Q3 2028) and provides 4× improvement with an associated multiplexed ground receiver; the 8-m ground aperture could be deployed later, without imposing requirements on the spacecraft, for another 4× improvement.